\documentclass[pra,twocolumn,showpacs,superscriptaddress]{revtex4}

\newcommand{\ket}[1]{| #1 \rangle}

\begin{document}

\title{Quantum polarization properties of two-mode energy eigenstates}

\author{Anita Sehat}
\affiliation{Department of Microelectronics and Information
Technology, Royal Institute of Technology (KTH), Electrum 229,
SE-164 40 Kista, Sweden}

\author{Jonas S\"{o}derholm}
\affiliation{Department of Microelectronics and Information
Technology, Royal Institute of Technology (KTH), Electrum 229,
SE-164 40 Kista, Sweden}

\affiliation{The Graduate University for Advanced Studies
(SOKENDAI), Hayama, Kanagawa 240-0193, Japan}

\affiliation{Institute of Quantum Science, Nihon University, 1-8
Kanda-Surugadai, Chiyoda-ku, Tokyo 101-8308, Japan}

\author{Gunnar~Bj\"{o}rk}
\email{gunnarb@imit.kth.se} \homepage{http://www.quantum.se}
\affiliation{Department of Microelectronics and Information
Technology, Royal Institute of Technology (KTH), Electrum 229,
SE-164 40 Kista, Sweden}

\author{Pedro~Espinoza}
\affiliation{Departamento de F\'{\i}sica, Universidad de
Guadalajara, Revoluci\'on 1500, 44420 Guadalajara, Jalisco,
M\'exico}

\author{Andrei B. Klimov}
\affiliation{Departamento de F\'{\i}sica, Universidad de
Guadalajara, Revoluci\'on 1500, 44420 Guadalajara, Jalisco,
M\'exico}

\author{Luis L. S\'{a}nchez-Soto}
\affiliation{Departamento de \'{O}ptica, Facultad de Ciencias
F\'{\i}sicas, Universidad Complutense, 28040 Madrid, Spain}

\date{\today}

\begin{abstract}
We show that any pure, two-mode, $N$-photon state with $N$ odd or
equal to two can be transformed into an orthogonal state using
only linear optics. According to a recently suggested definition
of polarization degree, this implies that all such states are
fully polarized. This is also found to be true for any pure,
two-mode, energy eigenstate belonging to a two-dimensional SU(2)
orbit. Complete two- and three-photon bases whose basis states are
related by only phase shifts or geometrical rotations are also
derived.
\end{abstract}

\pacs{42.25.Ja, 42.50.Dv, 42.50.Ar}

\maketitle

\section{Introduction}

The polarization of a propagating electromagnetic field is a
robust characteristic, which is relatively simple to manipulate
without inducing more than marginal losses. For this reason, many
recent experiments in quantum optics, such as Bell tests
\cite{Aspect,Kwiat}, quantum tomography \cite{Barbieri},
entanglement witnesses \cite{Bourennane}, quantum cryptography
\cite{Bennett,Muller}, and quantum dense coding \cite{Mattle},
have been performed using polarization bases.

As long as the polarization measurements involve only qubits,
encoded in two orthogonal polarization states of a single photon,
the classical theory of polarization \cite{Stokes} and the quantum
theory \cite{Collett} essentially coincide. However, for
multiphoton states, there is a divergence between the classical
and the quantum mechanical concepts of polarization. States that
have vanishing expectation values of all three Stokes parameters
are unpolarized according to the conventional classical theory,
but may result in full visibility in a quantum measurement
\cite{Prakash,Agarwal,Lehner,Usachev,OS}. Another example of
discrepancy between the classical and quantum notion of
polarization is the existence of states that are transformed into
orthogonal states by geometrical rotations of $\pm$60 degrees
around the propagation axis \cite{Tsegaye}. Classically, only
linearly polarized states can evolve into an orthogonal
polarization upon a rotation of $\pm$90 degrees. The apparent
``violations'' of the classical concept have led to the notion of
states with ``hidden polarization'' \cite{Klyshko}.

When discussing polarization properties of quantum states, it is
instructive to look back on the early discussion of unpolarized
light. In 1971, Prakash and Chandra \cite{Prakash} proposed that a
reasonable definition of an unpolarized state of light was to
require invariance of the state with respect to geometrical
rotations and phase shifts, or any combination thereof.
Restricting ourselves to the energy eigenstates, the corresponding
transformations form the group SU(2) \cite{Campos}. Following the
thread of Prakash and Chandra, and later Agarwal \cite{Agarwal}
and Lehner {\it et al.} \cite{Lehner}, we have proposed
\cite{SPIE} that the degree of polarization of a quantum state
should be given by the maximum observable generalized visibility
\cite{Bjork1} of the state under such SU(2) transformations. That
is, if it is possible to transform a state to an orthogonal state
by some combination of geometrical rotations and phase shifts,
then the state has a unit degree of quantum polarization.

Other attempts have been made to quantify quantum polarization.
One measure is due to Luis \cite{Luis}, where the degree of
polarization is expressed by the means of the dispersion of the
SU(2) $Q$-function over the Poincar\'{e} sphere. A quantity
$\Sigma$ that can be interpreted as the ``effective area'' of the
sphere where the $Q$-function is different from zero was defined,
and the smaller this area is, the higher the degree of
polarization. With this definition, SU(2) coherent states are
fully polarized, while the vacuum state, having an isotropic
$Q$-function, has zero degree of polarization. In contrast to our
measure, which only quantifies the smallest possible overlap
between the state and any rotated and phase-shifted state, Luis'
measure favors states that can become orthogonal (or almost
orthogonal) under the least ``action''. That is, a state whose
$Q$-function occupies only a small effective area on the
Poincar\'{e} sphere needs not be rotated by much before the
$Q$-functions of the original and the rotated states no longer
overlap. In Luis' theory, such states are assigned a large degree
of polarization. This difference between Luis' measure and ours
becomes poignant when we study the states $\ket{2,0}$ (two
horizontally, linearly polarized photons) and $\ket{1,1}$ (one
horizontally and one vertically, linearly polarized photon).
According to Luis' theory, for which the degree of polarization of
two-photon states can take on values between 0 and 4/9, these
states have the degree of polarization $4/9$ and $1/6$,
respectively. Our measure assigns the value unity for the degree
of polarization for both states, as will be shown below. The
reason is that by geometrical rotations only, both states may be
transformed into orthogonal states.

Another suggested definition of the degree of polarization for
multimode states was given by Karassiov {\it et al.}
\cite{Karassiov}. This measure is only comparable to ours in the
single-mode case (which treats two polarization modes). In this
regime, Karassiov's measure coincides with the ``classical''
definition of polarization based on the expectation values of the
Stokes operators. However, the objective of the present work is to
avoid the known problems associated with the classical definition.
For example, it has been experimentally demonstrated
\cite{Chekhova} that it is possible, by applying appropriately
chosen SU(2) transformations, to transform pure two-photon states
of various degree of classical polarization into an orthogonal
state.

In this paper, we use the definition of polarization degree given
in Ref.\ \cite{SPIE} and consider pure $N$-photon states. In order
to speak about polarization at all, we have assumed that we are
dealing with a propagating light field, for which we can define
two orthogonal transverse modes. Far from the source, all
electromagnetic fields propagating in isotropic media evolve
towards transverse fields. Expressing the polarization state in
terms of the excitation of these two modes is therefore
justifiable. In the following, we shall take these modes to be
plane-wave modes with the electric field directed in the
horizontal and the vertical directions. If the horizontal and
vertical modes have $m$ and $n$ photons respectively, we shall
denote this state as $\ket{m,n}$.

As phase shifts and geometrical rotations are lossless
transformations, our treatment based on these operations naturally
disintegrate into energy manifolds containing different number of
photons. Also from an experimentally point of view it is natural
to consider energy eigenstates, since photon counters are normally
used as detectors in experiments involving quantized polarization
states of light. Hence, the final projectors are photon number
states and therefore the polarization properties of two-mode
energy eigenstates are of significant current interest.

\section{Quantum degree of polarization}

Mathematically, differential phase shifts and geometrical
rotations can be easily expressed using the Stokes operators.
However, in order to make comparisons with work not related to
polarization easier, we will instead use the Schwinger boson
realization of the angular momentum operators \begin{eqnarray}
\hat{J}_x & = & \frac{\hat{S}_x}{2} = \frac{1}{2}(\hat{a}^\dagger \hat{b} + \hat{a} \hat{b}^\dagger) , \nonumber \\
\hat{J}_y & = & \frac{\hat{S}_y}{2} = \frac{1}{2 i}
(\hat{a}^\dagger \hat{b} - \hat{a} \hat{b}^\dagger) , \label{eq: first}\\
\hat{J}_z & = & \frac{\hat{S}_z}{2} = \frac{1}{2}(\hat{a}^\dagger
\hat{a} - \hat{b}^\dagger \hat{b}) , \nonumber \end{eqnarray}
which equal the corresponding Stokes operators divided by two as
indicated. Here $\hat{a}$ ($\hat{b}$) is the annihilation operator
of the horizontally (vertically) polarized mode \cite{YYY}. The
effects of a differential phase shift of $\alpha$ and a
geometrical rotation by $\theta/2$ are given by the operators
$e^{-i \alpha \hat{J}_z}$ and $e^{-i \theta \hat{J}_y}$,
respectively. Using these two physical operations, we can
construct unitary representations of the group SU(2) in the
different energy manifolds as \cite{Campos} \begin{equation}
\hat{U} (\beta,\theta,\alpha) = e^{-i \beta \hat{J}_z} e^{-i
\theta \hat{J}_y} e^{-i \alpha \hat{J}_z} . \label{eq:Tg}
\end{equation} Hence, any SU(2) transformation can be realized
using differential phase shifts and geometrical rotations alone,
and any such combination can be described by an operator of the
form (\ref{eq:Tg}).

Since these transformations preserve the total number of photons
$N$, we can treat the corresponding energy manifolds separately.
We thus use the fact that the Hilbert space of the two harmonic
oscillators can be expressed as a direct sum ${\cal H} \oplus_{N=0}^\infty {\cal H}_N$, where ${\cal H}_N$ is the Hilbert
space consisting of all two-mode $N$-photon states. Note that
${\cal H}_N$ has dimension $N+1$ and corresponds to a spin system
of spin $S = N/2$.

For two-mode bosonic states in the $N = 1$ manifold, there are
three mutually complementary operators $\hat{J}_z$, $\hat{J}_x$,
and $\hat{J}_y$, whose respective eigenstates have linear
polarization in directions $+$ and $\times$, and circular
polarization. Although the three Hermitian operators do not
commute, having the commutation relations $[ \hat{J}_x , \hat{J}_y
] = i \hat{J}_z$, where $x$, $y$, and $z$ can be cyclicly
permuted, they are mutually complementary (or unbiased)
\cite{Wootters} only in this manifold, which means that the
overlap between the corresponding normalized eigenstates is $1 /
\sqrt{2}$ in the Hilbert space of dimension $2$. The operators are
not complementary in manifolds for which $N \geq 2$.

A sensible approach to avoid the weaknesses of the definition
relying on the Stokes operators is to define a state that is not
invariant under all possible linear polarization transformations
to have a finite degree of quantum polarization. In an earlier
paper \cite{SPIE}, we have suggested a measure for the degree of
quantum polarization of two-mode states, based on this approach.
For pure states, the measure simplifies to \begin{equation}
\eta_{\rm q} = \sqrt{1- \min_{\beta,\theta,\alpha} | \langle \psi
| \hat{U} (\beta,\theta,\alpha) | \psi \rangle |^2} ,
\label{eq:PolDegDef}
\end{equation} where the overlap between the original state and
the transformed state is a measure of distinguishability between
the two states. According to this definition, any state that is
invariant under the SU(2) transformations $\hat{U}
(\beta,\theta,\alpha)$ is an unpolarized state and thus has zero
degree of quantum polarization. A fully polarized pure state, on
the other hand, satisfies
\begin{equation} \min_{\beta,\theta,\alpha} |\langle \psi| \hat{U}
(\beta,\theta,\alpha) |\psi\rangle | = 0 \quad \Leftrightarrow
\quad \eta_{\rm q}=1 . \label{eq:UnitDeg} \end{equation} As we
shall show below, for any pure $N$-photon state with $N$ odd or $N
= 2$, there exists a transformation of the form $\hat{U}
(\beta,\theta,\alpha)$ that transforms the state into an
orthogonal one. That is, any pure state in these manifolds is
fully polarized.

The general form of an unpolarized quantum state was derived
already in the work of Prakash and Chandra \cite{Prakash} (see
also \cite{Agarwal,Lehner,OS}). The only unpolarized $N$-photon
state is \begin{equation} \hat{\rho} = \frac{1}{N+1} \sum_{n=0}^N
| n,N-n \rangle \langle n,N-n | , \end{equation} which is a
maximally mixed state. In other words, it is the $N$-photon state
with the largest von Neumann entropy.

\section{SU(2) orbits}
\label{group orbits}

From our definition of degree of quantum polarization, it is clear
that all states that can be transformed into each other by an
operator of the form (\ref{eq:Tg}) have the same degree of quantum
polarization. Such a set of $N$-photon states form an {\em orbit}
(of the SU(2) group), and $\mathcal{H}_N$ is a union of disjoint
orbits.

The case of $N=1$ (a polarization qubit) is trivial: there is only
a single orbit. It is characterized by two real parameters, which
covers the whole two-dimensional space, i.e., the Poincar\'{e}
sphere. As discussed above, the classical theory of polarization
coincides with the quantum theory for this case. We know from the
classical theory that we can always find a combination of
geometrical rotations and phase shifts that transforms a point on
the Poincar\'{e} sphere to a diametrically opposite point,
implying an orthogonal polarization. Hence, all pure polarization
qubit states have a unit degree of quantum polarization.

For $N>1$ there are two types of orbits \cite{Nuno}:

\begin{enumerate}
\item Orbits of states with nontrivial stability group U(1) and
any additional group of discrete symmetry. These orbits are
two-dimensional and are generated from the bare basis states by
applying the operator of the group representation. However, due to
the relation \cite{Rotation}
\begin{equation}
\hat{U} \left( 0,\pi,0 \right) |n,N-n \rangle = (-1)^{n} |N-n,n
\rangle , \label{eq:reflection}
\end{equation}
there are only $\lfloor N/2 \rfloor + 1$ orbits of this type,
where $\lfloor x \rfloor$ denotes the largest integer smaller than
or equal to $x$. These orbits are isomorphic to $\mathcal{S}_2/H$,
where $\mathcal{S}_2$ denotes the two-dimensional sphere and $H$
is the discrete  group of symmetry.

\item Orbits which allow only groups of discrete symmetries. These
orbits are three-dimensional and isomorphic to $\mathcal{S}_3/H$.
The orbit space is defined as the quotient $\mathcal{H}_N$/SU(2)
and is $(2N-3)$ dimensional for $N > 1$. For example, for $N = 2$,
we have dim~[$\mathcal{H}_2$/SU(2)] $= 1$, which means that a
single (real) parameter is needed to separate the orbits. However,
for $N = 3$, we have dim~[$\mathcal{H}_3$/SU(2)] $= 3$ and one
hence needs 3 parameters.
\end{enumerate}

\section{Orbits of type 1}
\label{sec-orbits type 1}

Let us first consider the orbits of type 1. These orbits can be
labeled by the value $n \in \{ 0,1,\ldots,\lfloor N/2 \rfloor \}$.
From Eq.~(\ref{eq:reflection}), it is clear that for every state
belonging to an orbit with $n \neq N/2$, there exists an
orthogonal state within the same orbit, because the states $|n,N-n
\rangle$ and $|N-n,n \rangle$ are then orthogonal. Therefore, all
states belonging to orbits with $n \neq N/2$ are fully polarized.

States belonging to orbits for which $N$ is an even number and $n
= N/2$, also has unit quantum polarization. To prove this, we note
that \begin{equation} \langle N/2,N/2 | \hat{U} (0,\theta,0) |
N/2,N/2 \rangle = P_{N/2}(\cos \theta), \label{eq:rN}
\end{equation} where $P_n(x)$ is the Legendre polynomial of order
$n$ as a function of $x$. Such a polynomial has $n$ zeros in the
interval $-1<x<1$, so it is always possible to find a solution to
the equation $\langle N/2,N/2 | \hat{U} (0,\theta,0) | N/2,N/2
\rangle = 0$.

In  conclusion, we have shown that all pure states belonging to
orbits of type 1 have unit degree of quantum polarization
irrespective of their excitation.

\section{Two-photon orbits of type 2}

It is fairly easy to show that all pure two-photon states ($N=2$)
have unit degree of quantum polarization. Defining the step
operators $\hat{J}_\pm = \hat{J}_x \pm i \hat{J}_y$, transitions
between any pair of orbits can be realized by unitary operators of
the form
\begin{equation}
e^{\vartheta \left( \hat{J}_-^2 - \hat{J}_+^2 \right)/2} = \left[
\matrix{\cos  \vartheta & 0 & \sin \vartheta  \cr 0 & 1 & 0 \cr
-\sin  \vartheta & 0 & \cos  \vartheta} \right] ,
\label{eq:TransitionOperator} \end{equation} which correspond to
rotation matrices. Here and throughout the paper, we have used the
basis ($\ket{0,N},\ket{1,N-1},\ldots,\ket{N,0}$) when writing
vectors and matrices. All SU(2) orbits can now be generated by
applying the operators (\ref{eq:TransitionOperator}) to the state
$\ket{0,2}$ or $\ket{2,0}$ (but not $\ket{1,1}$). Defining the
states \begin{equation} \ket{\psi(\vartheta)} = e^{\vartheta
\left( \hat{J}_-^2 - \hat{J}_+^2 \right)/2} \ket{2,0} = \sin
\vartheta \ket{0,2} + \cos \vartheta \ket{2,0} , \label{eq:psi2}
\end{equation} the orbits can be identified by a single parameter, which is
in agreement with our findings in Sec.\ \ref{group orbits}. Due to
the symmetry expressed in Eq.\ (\ref{eq:reflection}), it suffices
\cite{Nuno} to consider $0 \leq \vartheta \leq \pi/4$. We note
that the end points $\vartheta = 0$ and $\vartheta = \pi/4$
correspond to the states $|2, 0 \rangle$ and $(\ket{0,2} +
\ket{2,0})/\sqrt{2}$, which belong to the two orbits of type 1
characterized by the values $n = 0$ and $n = 1$ in Eq.\
(\ref{eq:reflection}), respectively. The latter can be seen from
the equality
\begin{equation} \hat{U} (\beta,\pm \pi/2,\mp \pi/2) \ket{\psi
(\pi/4)} = i \ket{1,1} . \label{eq:PsiPiOver4Orbit}
\end{equation}

In order for the state $\ket{\psi (\vartheta)}$ to have unit
degree of polarization (\ref{eq:UnitDeg}), we must have
\begin{eqnarray} \lefteqn{\langle \psi (\vartheta) | \hat{U}
(\beta,\theta,\alpha)
\ket{\psi(\vartheta)} = \cos (\alpha - \beta) \sin 2 \vartheta \sin^2 (\theta/2)} & & \nonumber \\
& & + [\cos (\alpha + \beta) - i \sin (\alpha + \beta) \cos 2
\vartheta] \cos^2 (\theta/2) = 0 . \quad \quad \end{eqnarray}
Eight solutions that are independent of $\vartheta$ are easily
found. They are given by $\theta = \pi$, $\beta = \alpha \pm
\pi/2$, and $\alpha = \pm \pi/4$ or $\alpha = \pm 3 \pi/4$. All
these solutions give the same physical final state, described by
$\cos \vartheta \ket{0,2} - \sin \vartheta \ket{2,0}$.

Since any pure two-photon state can be obtained by applying an
SU(2) operator to some orbit-generating state
$\ket{\psi(\vartheta)}$, we conclude that all pure two-photon
states can be mapped onto an orthogonal one using only linear
optics. According to our definition (\ref{eq:PolDegDef}), these
states thus have unit degree of polarization.

\section{Complete two-photon bases}
\label{sec:TwoPhotonBases}

As we have already noted, the state \begin{equation} \ket{\psi
(\pi/4)} = \frac{1}{\sqrt{2}} (\ket{0,2} + \ket{2,0})
\label{eq:PsiPiOver4} \end{equation} belongs to one of the two
orbits of type 1 for $N = 2$. It can also be seen as a peculiar
circularly polarized state \cite{Tsegaye}. As a consequence of its
circular nature, it can be transformed into an orthogonal state
according to
\begin{equation} \hat{U} (\pm \pi/2,\theta,0) \ket{\psi(\pi/4)} = \pm \frac{i}{\sqrt{2}}(\ket{0,2} - \ket{2,0})
\label{eq:PsiPiOver4theta} \end{equation} for any value of
$\theta$. According to Eq.\ (\ref{eq:PsiPiOver4Orbit}), SU(2)
operators can also transform $\ket{\psi (\pi/4)}$ into a state
that is orthogonal to both (\ref{eq:PsiPiOver4}) and
(\ref{eq:PsiPiOver4theta}). Hence, in this orbit of type 1, the
SU(2) transformations generate the whole basis set. We have
already used this fact to experimentally generate basis states
that differ only by phase shifts \cite{Trifonov} or geometrical
rotations \cite{Tsegaye}. That the states
(\ref{eq:PsiPiOver4Orbit}), (\ref{eq:PsiPiOver4}), and
(\ref{eq:PsiPiOver4theta}) are orthogonal was recently pointed out
by Chekhova \textit{et al.} \cite{Chekhova}, who denoted them
$\ket{HV}$, $\ket{RL}$, and $\ket{D \overline{D}}$, respectively,
reflecting the fact that they represent one photon in each of a
horizontal-vertical linear basis, the right- and left-hand
circularly polarized basis, and the linear basis $\pm 45$ degrees
from the vertical. That is, they are the eigenstates with
eigenvalue zero of the operators $\hat{J}_z$, $\hat{J}_y$, and
$\hat{J}_x$, respectively.

States that can generate whole basis sets using only linear
transformations are very useful, since these transformations are
easily realized experimentally. In particular, it is desirable to
find states that can generate a whole basis set by using only
phase shifts or geometrical rotations. As phase shifts and
geometrical rotations are described by the operators $e^{-i \alpha
\hat{J}_z}$ and $e^{-i \theta \hat{J}_y}$, such bases can be
generated from equipartition states in the $\hat{J}_z$- and
$\hat{J}_y$-basis \cite{PhysScr}, respectively.

An equipartition state in the $\hat{J}_z$-basis, i.e., the
horizontal-vertical basis, can be realized by an SU(2)
transformation acting on the state $|\psi (\pi/4) \rangle$ since
\begin{equation} |\xi_1 \rangle = \frac{\hat{U}
(0,\theta_z,\pi/2)}{\sqrt{2}} \left[ \matrix{1 \cr 0 \cr 1}
\right]= \frac{i}{\sqrt{3}} \left[ \matrix{1 \cr -1 \cr -1}
\right] , \end{equation} where $\theta_z = \pi/2-
\arccos(1/\sqrt{3})$. This state then forms a complete basis
together with the states generated by subsequent phase shifts of
$2 \pi/3$ and $-2 \pi/3$
\begin{equation} |\xi_{2,3} \rangle = \hat{U} (0, 0, \pm  2 \pi/3) |\xi_1
\rangle = \frac{i}{\sqrt{3}} \left[ \matrix{e^{\pm i 2 \pi/3} \cr
-1 \cr -e^{\mp i 2 \pi/3}} \right] . \end{equation}

Starting with the same state $|\psi (\pi/4) \rangle$, an
equipartition state in the $\hat{J}_y$-basis, i.e., the circularly
polarized basis, can be obtained by application of the phase shift
$\alpha_y = \arctan(1/\sqrt{2}) - \pi/2$. Expressed in the
horizontal-vertical basis, we then have \begin{equation} |\psi_1
\rangle = \frac{\hat{U} (0,0,\alpha_y)}{\sqrt{2}} \left[ \matrix{1
\cr 0 \cr 1} \right]= \frac{1}{\sqrt{6}} \left[ \matrix{1 - i
\sqrt{2} \cr 0 \cr 1 + i \sqrt{2}} \right] . \end{equation} This
state can be transformed into two other orthogonal two-mode states
under geometrical rotations of $\pm 60$ degrees \begin{equation}
|\psi_{2,3} \rangle = \hat{U} (0,\pm 2 \pi/3, 0) |\psi_1 \rangle \frac{1}{2 \sqrt{3}} \left[ \matrix{\sqrt{2} + i \cr \pm i
\sqrt{6} \cr \sqrt{2} - i} \right] , \end{equation} which together
with $|\psi_1 \rangle$ form a complete orthonormal basis.

Since any unitary transformation $\hat{V}$ preserves the inner
products, new bases can be created by applying $\hat{V}$ to the
original basis states. If the original basis states belong to the
same orbit and we use an SU(2) transformation, all basis states
remain within the orbit. Any state belonging to such an orbit can
thus be made a basis state of a complete basis by an appropriately
chosen SU(2) transformation.

The orbit considered above, which is characterized by $\vartheta \pi/4$, is easily reached experimentally using a photon pair
generated by spontaneous parametric down-conversion. The fact that
this orbit spans the whole Hilbert space ${\cal H}_2$ has been
exploited in the experimental realization of relative-phase states
\cite{Trifonov}, three mutually orthogonal polarization states
\cite{Tsegaye}, and two-mode, two-photon qutrits
\cite{Burlakov,Bogdanov}.

\section{Pure states with odd number of photons}

Let us denote an arbitrary pure $N$-photon state as
\begin{equation} | \chi \rangle = \sum_{n=0}^N r_n e^{i \varphi_n}
\ket{n,N-n} . \end{equation} For $N$ odd, we then find
\begin{eqnarray} \lefteqn{\langle \chi | \hat{U}
(0,\pi,\alpha) | \chi \rangle =} & & \nonumber \\
& & -i 2 \sum_{k=0}^{\frac{N+1}{2}} (-1)^k r_k r_{N-k} \sin \left(
\varphi_k - \varphi_{N-k} + \frac{(N - 2 k) \alpha}{2} \right) .
\nonumber \\ \label{eq:Nodd} \end{eqnarray} We thus have $\langle
\chi | \hat{U} (0,\pi,0) | \chi \rangle = - \langle \chi | \hat{U}
(0,\pi,2 \pi) | \chi \rangle$. Since the expression
(\ref{eq:Nodd}) is purely imaginary, there exists at least one
value of $\alpha$ for which a state orthogonal to $| \chi \rangle$
is obtained. Hence, all pure states with a given odd number of
photons have unit degree of quantum polarization.

We note that for any $N$-photon state, a differential phase shift
of $2 \pi$ does not change the physical state. However, for odd
$N$, it introduces an overall phase factor that equals -1, which
we made use of in the proof above.

\section{Complete three-photon bases}

Noting that the state $\ket{\psi (\pi/4)}$, which we used to
generate complete two-photon bases in Sec.\
\ref{sec:TwoPhotonBases}, is invariant under interchange of the
horizontally and vertically polarized modes, let us now consider
the three-photon states \begin{eqnarray} \ket{\zeta_1} & = & \frac{1}{\sqrt{2}} (\ket{0,3} + \ket{3,0}) , \\
\ket{\zeta_2} & = & \frac{1}{\sqrt{2}} (\ket{1,2} + \ket{2,1}) .
\end{eqnarray} Like $\ket{\psi (\pi/4)}$, these states are symmetric with
respect to the horizontal and vertical modes, however they belong
to orbits of type 2. Application of the transformation $\hat{U}
(0,\theta,\pi/2)$ to $\ket{\zeta_1}$ gives \begin{equation}
\frac{\hat{U} (0,\theta,\pi/2)}{\sqrt{2}} \left[ \matrix{1 \cr 0
\cr 0 \cr 1} \right] = \frac{e^{i 3 \pi/4}}{\sqrt{2}} \left [
\matrix{u_{11} + i u_{14} \cr - u_{12} + i u_{13} \cr u_{13} + i
u_{12} \cr - u_{14} + i u_{11}} \right] , \end{equation} where
$u_{ij}$ are the real matrix elements of the unitary operator
$e^{-i \theta \hat{J}_y}$. In order to generate a complete basis
in manifold $N = 3$, we now look for an equipartition state in
analogy with the treatment in Sec.\ \ref{sec:TwoPhotonBases}. That
is, we want the magnitude of all the probability amplitudes to
equal $1/\sqrt{N+1}=1/2$, which implies $u_{11}^{2} + u_{14}^{2} u_{12}^{2} + u_{13}^{2} = 1/2$. This requirement can be fulfilled
by choosing one of the rotation angles $\theta_\pm = \arccos (\pm
1/\sqrt{3})$, which give
\begin{equation} \frac{\hat{U} (0,\theta_\pm,\pi/2)}{\sqrt{2}}
\left[ \matrix{1 \cr 0 \cr 0 \cr 1} \right] = \frac{\pm i e^{\pm i
\frac{3}{4} \arccos \frac{1}{3}}}{2} \left[ \matrix{1 \cr \frac{i
\mp \sqrt{2}}{\sqrt{3}} \cr \frac{1 \pm i \sqrt{2}}{\sqrt{3}} \cr
i} \right] . \end{equation} In fact, an equipartition state in the
horizontal-vertical basis can be created by applying any of the
eight operators $\hat{U} (0,\pm \theta_\pm,\pm \pi/2)$ to the
state $\ket{\zeta_1}$ or $\ket{\zeta_2}$. For example, we have
\begin{equation} \frac{\hat{U} (0,\theta_\pm,\pi/2)}{\sqrt{2}} \left[
\matrix{0 \cr 1 \cr 1 \cr 0} \right] = \frac{e^{\pm i \frac{1}{4}
\arccos \frac{1}{3}}}{2} \left[ \matrix{1 \cr \frac{-i \pm
\sqrt{2}}{\sqrt{3}} \cr \frac{-1 \mp i \sqrt{2}}{\sqrt{3}} \cr i}
\right] . \end{equation} Subsequent phase shifts of $\pi/2$,
$\pi$, and $3 \pi/2$ to the equipartition states will then
generate complete bases. That is, for any equipartition state
$\ket{\epsilon_0}$, the states \begin{equation} \ket{\epsilon_k} \hat{U} (0,0,k \pi/2) \ket{\epsilon_0} , \quad k = 0,1,2,3,
\end{equation} are mutually orthonormal.

By applying a geometrical rotation of 45 degrees followed by a
phase shift of $\beta_y = \arccos (-\sqrt{2/3})$ to the two
symmetrical states $\ket{\zeta_1}$ and $\ket{\zeta_2}$, one can
also obtain equipartition states in the circularly polarized
basis. In the horizontal-vertical basis, we then have
\begin{equation} \frac{\hat{U} (\beta_y,\pi/2,0)}{\sqrt{2}} \left[
\matrix{1 \cr 0 \cr 0 \cr 1} \right] = \frac{-i e^{-i \frac{3}{4}
\arccos \frac{1}{3}}}{2} \left[ \matrix{1 \cr 0 \cr \frac{1 + i 2
\sqrt{2}}{\sqrt{3}} \cr 0} \right] \end{equation} and
\begin{equation} \frac{\hat{U} (\beta_y,\pi/2,0)}{\sqrt{2}} \left[
\matrix{0 \cr 1 \cr 1 \cr 0} \right] = \frac{-i e^{-i \frac{3}{4}
\arccos \frac{1}{3}}}{2} \left[ \matrix{\sqrt{3} \cr 0 \cr
-\frac{1 + i 2 \sqrt{2}}{3} \cr 0} \right] . \end{equation} Each
of these states form a complete basis together with the states
obtained by geometrically rotating the respective state by 45, 90,
and 135 degrees. The corresponding transformations are given by
the operators $\hat{U} (0, k \pi/2, 0)$, where $k = 1,2,3$.

Also in the case of three photons, the orbits to which some
particular symmetrical states belong thus span the whole Hilbert
space, and allow complete bases to be generated by applying only
phase shifts or geometrical rotations.

\section{Discussion and conclusions}

Polarization properties of quantum states deviate from the
properties one may extrapolate from classical physics. The
fundamental reason is the concept of orthogonality, which, for
classical states, takes on a direct geometrical meaning (in the
plane perpendicular to the propagation direction), whereas
orthogonality in Hilbert spaces of dimension greater than 2
instead implies distinguishability. This means that in Hilbert
spaces of dimension greater than 2, there are more than two
orthogonal states of polarization, as shown above.

The spaces spanned by two- and three-photon states are still
tractable, and we have shown that all pure states in these spaces
have unit degree of quantum polarization. This was also found to
be true for all pure states with any given odd number of photons.
That is, by using only geometrical rotations and phase shifts it
is always possible to transform any such state into an orthogonal
one. Equivalently, using the proper observable, the transformation
will result in a unit-visibility projection probability. We have
also shown that there exist complete two- and three-photon bases
whose basis states are related by transformations that can be
realized using linear optics. In particular, we derived such bases
with basis states related by only phase shifts or geometrical
rotations. For two-photon states, these properties have already
been exploited in various applications of quantum optics.

Finally, we note that it may be possible to generalize several of
the results presented here. For example, it is natural to ask if
all pure, two-mode, energy eigenstates can be made orthogonal
using linear optics. It would also be interesting to know which
SU(2) orbits can be used to generate complete bases.

\begin{acknowledgments}
This work was supported by INTAS grant 211-2122, the Swedish
Research Council (VR), and the Swedish Foundation for Strategic
Research (SSF). One of the authors (GB) acknowledges stimulating
discussions with Professors S.~Kulik and M. Chekhova.
\end{acknowledgments}


\begin{thebibliography}{99}
\bibitem{YYY} Actually, for the rotation to be positive with
respect to the direction of propagation of the field, the positive
field direction of the mode with the associated annihilation
operator $\hat{a}$, the corresponding direction associated
annihilation operator $\hat{b}$, and the propagation direction
must form a right-hand system.
\bibitem{Rotation} Physically, the transformation $\hat{U}
\left( 0,\pi,0 \right)$ corresponds to a geometrical rotation
about the $z$-axis, the axis of propagation, by $\pi/2$. That is,
the two linearly polarized modes are interchanged.
\end{thebibliography}
\end{document}